 \documentclass[final,12p,times,twocolumn]{elsarticle}

\usepackage{amssymb}
\usepackage{textgreek}
\usepackage{textcomp}
\usepackage{mathtools}
\usepackage{url}
\usepackage{cite}
\usepackage{xcolor}
\usepackage{tabularx}
\usepackage{enumitem}
\usepackage{hhline}

\journal{Nucl. Instr. Meth. Phys. Res.}

\begin{document}


\begin{frontmatter}



\title{A Double-Layered Water Cherenkov Detector Array for Gamma-Ray Astronomy
}


\author[inst1]{Samridha Kunwar\corref{cor1}}
\author[inst1]{Hazal Goksu\corref{cor2}}
\author[inst1]{Jim Hinton}
\author[inst2]{Harm Schoorlemmer}
\author[inst3]{Andrew Smith}
\author[inst1]{Werner Hofmann}
\author[inst1]{Felix Werner}

\affiliation[inst1]{organization={Max-Planck-Institut für Kernphysik},
            addressline={Saupfercheckweg 1}, 
            city={Heidelberg},
            postcode={69117}, 
            country={Germany}}

\affiliation[inst3]{organization={Department of Physics, University of Maryland},
            city={College Park},
            postcode={20742}, 
            state={Maryland},
            country={USA}}
            
\affiliation[inst2]{organization={IMAPP, Radboud University Nijmegen},
            city={Nijmegen},
            country={The Netherlands}}
            
\cortext[cor1]{email : samridha.kunwar@mpi-hd.mpg.de}
\cortext[cor2]{email : hazal.goksu@mpi-hd.mpg.de}

\begin{abstract}
Ground-level particle detection is now a  well-established approach to TeV $\gamma$-ray astronomy. Detection of Cherenkov light produced in water-filled detection units is a proven and cost-effective method. Here we discuss the optimization of the units towards the future Southern Wide-field Gamma-ray Observatory (SWGO). In this context, we investigate a new type of configuration in which each water Cherenkov detector (WCD) unit in the array comprises two chambers with black or reflective walls and a single photomultiplier tube (PMT) in each chamber. We find that this is a cost effective approach that improves the performance of the WCD array with respect to current approaches. A shallow lower chamber with a PMT facing downwards enables muon tagging and the identification of hadron-induced air showers, which are the primary source of background in $\gamma$-ray astronomy. We  investigate how $\gamma$/hadron separation power and achievable angular resolution depend on the geometry and wall reflectivity of the detector units in this configuration. We find that excellent angular resolution, background rejection power and low-energy response are achievable in this double-layer configuration, with the aid of reflective surfaces in both chambers.
\end{abstract}


\end{frontmatter}


\section{Introduction}
\label{intro}
Ground-level particle-based detection of air showers is a rapidly developing approach to $\gamma$-ray astronomy at very high energies, with cosmic-ray protons and nuclei as the main source of background. High-density arrays maximise the number of detectable particles per air shower. Water Cherenkov detectors (WCD) are water-filled detection units that detect Cherenkov light produced by air showers reaching ground level and have been proven to be an effective way of achieving large array area, as demonstrated by HAWC (High-Altitude Water Cherenkov)~\citep{Abeysekara_2017} and LHAASO (Large High Altitude Air Shower Observatory) coverage~\citep{Aharonian_2021}. The Southern Wide-field Gamma-ray Observatory (SWGO) \citep{Abreu_2019, Hinton_2021}, is a project towards constructing a large detector array in the southern hemisphere with an advanced detector design and superior performance compared to both HAWC and LHAASO. The performance of WCD arrays is largely driven by high altitude~\citep{Sinnis_2007}, large array area and large fill-factor~\citep{Schoorlemmer_2019}, but the particle detection thresholds of the individual WCD units and their response characteristics will influence the threshold energy and performance of the array (see e.g.~\citep{Assis_2017}). This paper investigates the reference design for SWGO; a double-layered WCD array. Several other advanced designs such as a shallow WCD with 4 photomultiplier tubes (PMTs) are also being considered~\citep{Conceicao_2021}. 

The role of an individual detector unit in a ground-particle-based $\gamma$-ray instrument is to measure the local shower particle number or energy density, assign a local arrival time and ideally provide  information for $\gamma$/hadron separation.
Muon identification, as a means of hadronic background rejection (see, e.g.~\citep{Schoorlemmer_2019}), can be implemented either using separate detector elements, such as the LHAASO buried muon detectors~\citep{Aharonian_2021} or by discrimination within a standard WCD unit. Additionally, the topology of the shower amplitude distribution such as the charge distribution close to the shower core and the clustering characteristics of hits far from the core region also provide information for effective $\gamma$/hadron discrimination as demonstrated by Milagro~\citep{Atkins_2003}.

The two-chamber concept, comprising a water volume separated into optically isolated top and bottom chambers, explored here, provides a (potentially) cost-effective general-purpose element with reasonable time and amplitude resolution as well as muon identification~\citep{Letessier_2014}. The principle is well established: most shower muons will pass through the entire detector element with only ionisation losses and produce Cherenkov light in both chambers. Electrons and $\gamma$-rays generate cascades that penetrate to the lower chamber only in the case of high energy initial particles, and in any case, the comparison of the signals in both sections allows discrimination from a through-going muon. As the lower chamber captures the cascade products of only high energy shower particles, it can also extend the detector dynamic range close to the shower core in the case of high energy primary $\gamma$-rays. 

The quantities to be optimised for the double-layered WCD unit are the dimensions of the two chambers, the reflective properties of the internal surfaces and the photosensors. The main current instruments of this type, HAWC and the LHAASO~\citep{Aharonian_2021}, employ non-reflective surfaces for detector water volumes made of single chambers, with a diameter-to-depth aspect ratio of 1.6--1.8. 

We first discuss particle detection efficiency, photon timing distributions and background trigger rates of individual double-layered WCD units and then discuss the performance of the WCD array in its entirety. After a brief overview of the double-layered WCD design (Section~\ref{sec: design overview}) and of the simulation tools used (Section~\ref{sec: simulation overview}), we explore the trade-offs associated with the reflectivity of the unit surfaces and the overall unit dimensions (Sections \ref{sec:upper} and \ref{sec:lower}). In these sections we also compare our results from the double-layered WCDs with single-chamber WCDs similar to HAWC and LHAASO units.
In a second step we then address the energy threshold, $\gamma$/hadron separation power and angular resolution of an ensemble of detector units (Section \ref{sec:array}). 

\section{Design Overview}
\label{sec: design overview}

A candidate WCD design for SWGO with muon separation potential is a double-layered unit, illustrated in Fig.~\ref{fig:unit}, that comprises the following building blocks:

\begin{figure*}
  \includegraphics[width=1.\textwidth]{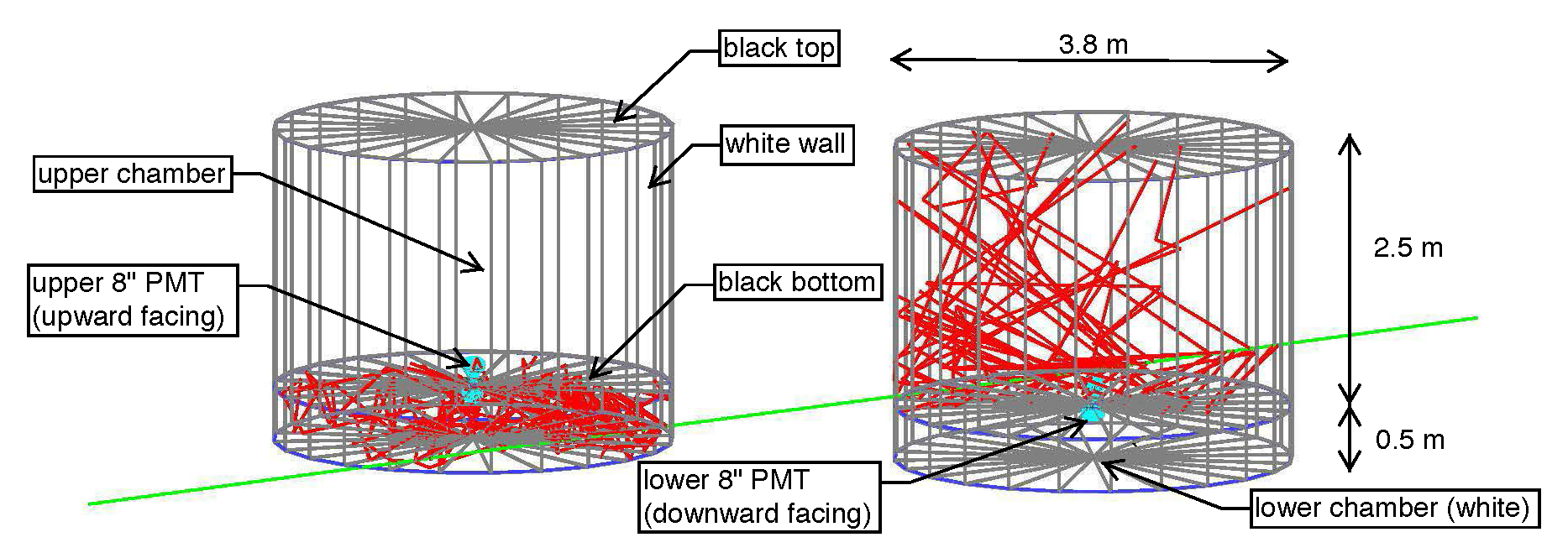}
\caption{Cylindrical double-layered WCD design comprising an upper chamber of 3.8\,m diameter and 2.5\,m depth, here with white walls and black bases (top and bottom) and an entirely white lower chamber of 0.5\,m depth. The upper chamber contains an 8-inch PMT facing upwards, and the lower chamber has an 8-inch PMT facing downwards. For illustration, a simulated muon (green track) is shown that passes through both units and produces Cherenkov photons (red tracks).}
\label{fig:unit}       
\end{figure*}

\begin{itemize}

\item Upper Chamber:
A light-tight chamber with a lining that may be black or reflective and a single upward-facing PMT. This chamber provides timing information and an estimate of total local particle energy per unit area. The upward-facing PMT ensures that non-reflected Cherenkov photons with the smallest time dispersion are detected first.

\item Lower Chamber:
A light-tight chamber with a highly reflective lining, containing a single PMT facing downwards (for improved uniformity of  response). This chamber enables muon tagging as only a small fraction of the higher energy photons and electrons at ground level can punch through into the lower chamber, while nearly all muons will pass through the entire detector unit as the mean linear stopping power ($\rho  \langle  - dE/dx \rangle$) for muons  in water is only 2 MeV/cm \citep{2001ADNDT..78..183G}.
\end{itemize} 

The two PMTs are connected to each other with a PMT support such that one faces upwards in the upper chamber, and the other faces downwards in the lower chamber.
\noindent
The reference design uses a 3.8 m diameter tank, motivated by the relative ease of road transportation of pre-fabricated (e.g. rotomolded) units up to this size. Alternative designs with different diameters are possible and included in our studies for single tank simulations (see Section \ref{sec:upper}). The depths of the two chambers and their reflective properties are investigated in Sections \ref{sec:upper} and \ref{sec:lower}. 

\section{Overview of Simulation Tools}
\label{sec: simulation overview}

To simulate air showers, we use the CORSIKA~7.7400 simulation package~\citep{Heck_1998}. For the standard simulated event set, we select QGSJet-II.04~\citep{Ostapchenko_2011} for energies above 80 GeV. UrQMD 1.3.1~\citep{Bass_1998, Bleicher_1999} treats the low energy hadronic interactions and for electromagnetic processes, we use the EGS4 electromagnetic model~\citep{Nelson_1990}.

We use GEANT4~\citep{GEANT4_2002} within a simulation framework adapted from that of the HAWC collaboration, to simulate the WCD response to the secondary air shower particles from 20 m above the detector unit.  This simulation framework, called \textit{HAWCSim}, has been extensively used for studies related to HAWC and has been verified by the HAWC Collaboration ~\citep{Springer_2017, ABEYSEKARA2015125, Torres_2021}. 

The UNIFIED~\citep{Unified_1996} model in GEANT4 is adapted to describe the reflectivity of materials such as Polypropylene (low reflectivity - 10\% at 450 nm, from now on referred to as `black') as used by HAWC and those with a rough surface such as Tyvek (high reflectivity - 92\% at 450 nm, which from now on we refer to as `white')~\citep{Nozka_2011} as used by the Pierre Auger Observatory~\citep{Allekotte_2008}. The standard deviation of the distribution of the micro-facet orientations (taken for rough surfaces), is set as $\sigma_\alpha = 0.17$ rad for the Tyvek surface. 

\par A nominal water absorption length of 17\,m at 400\,nm is used in the simulations. This water absorption length is reasonable when compared with the numbers measured for the current detector arrays. The LHAASO collaboration purifies its water to get an absorption length longer than 15\,m for around 400\,nm \citep{Ma_2022}, meanwhile studies performed by the HAWC collaboration show attenuation lengths  varying between 5\,m and 16\,m for 405\,nm~\citep{Abeysekara_2015}. 

We model the PMT in the simulations after the 8-inch HPK R5912-20 with a photo-cathode quantum efficiency of 20\% at 450 nm~\citep{Hamamatsu_2019}. The central 10'' PMT for one of the configurations in our comparisons (configuration B from Section \ref{sec: sensitivity}, similar to the HAWC main array tanks) is modeled after the 10'' R7081 with a photo-cathode efficiency of 30\% \citep{Hamamatsu_2019}.

\section{Upper Chamber Optimisation}
\label{sec:upper}

For $\gamma$-ray induced extensive air showers (EAS) at typical detector altitudes, the energy distributions of secondaries (in terms of number per log energy interval) varies with particle type ($dN/dlogE$). This number density peaks around $\sim$6\,MeV for secondary photons,  $\sim$20 \,MeV for electrons and 2--3\,GeV for muons~\citep{Schoorlemmer_2019}. A high detection probability for particles of these energies is  desirable for triggering  and reconstructing showers, combined with a precise determination of particle arrival time. Below we discuss the impact of geometry and material properties on the performance of the upper chamber, concerning particle detection efficiency (Section\,\ref{sec: sensitivity}) and arrival time measurement (Section\,\ref{sec: timing}). Chamber characteristics will also influence the background trigger rates of each WCD unit (Section\,\ref{sec: noise rates}), that in turn determine the trigger condition for the array and hence its energy threshold. 

 We study single unit simulations in this section, however the performance of the array of WCD units as a whole is ultimately the guiding factor for upper chamber geometry and material optimization. Array simulations to optimize the upper chamber are discussed in Section \ref{sec:array}. 

\subsection{Particle Detection Efficiency}
\label{sec: sensitivity}

 The depth of the chamber must be at least several radiation lengths ($\sim$4--5\,$X_0 \approx 1.5$--$1.8$\,m) for calorimetric detection of electromagnetic shower particles. The opening angle of the Cherenkov cone in water is $\sim$41\textdegree, and the pair production length of high energy $\gamma$-rays is $\frac{9}{7} X_0 $, where $X_0$ is the radiation length, corresponding to $\sim$46\,cm in water. The diameter of the chamber, along with the depth, determines the probability of collecting Cherenkov photons at the PMT, as prompt photons or after some number of diffuse reflections. As stated in Section \ref{sec: design overview}, the diameter for the reference design is fixed at 3.8\,m for logistical reasons, however in case of alternative designs for SWGO or any future WCD array, other diameters should be possible.

\begin{figure}[!ht]
\centering
\includegraphics[width=0.48\textwidth]{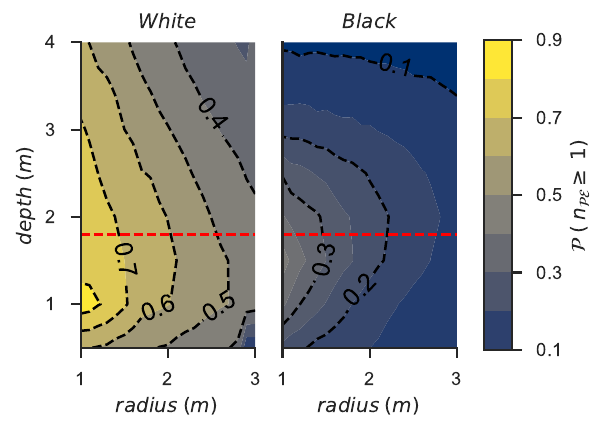}
\caption{Probability of detection of one or more photo-electrons as a function of upper chamber radius and depth, for vertical 10\,MeV $\gamma$-rays injected across the top surface of the chamber, for a chamber with entirely white walls (left) and a chamber with entirely black walls (right). A dashed horizontal red line shows the depth corresponding to $\approx$5 radiation lengths. The reference design for SWGO has a radius of 1.91\,m (3.8\,m in diameter). }
\label{fig:dim_scan_10MeV}
\end{figure}

Noting that prompt Cherenkov photons are important for timing (see Section~\ref{sec: timing}) and the minimum depth criteria explained above, we investigate the optimum chamber aspect ratio. To illustrate the dependencies, 
we inject vertical 10\,MeV $\gamma$-rays uniformly across the top surface of a double-layered WCD unit with an 8-inch PMT centered at the bottom. Fig.~\ref{fig:dim_scan_10MeV} shows the probability of detecting at least one Cherenkov photon, as a function of chamber radius and depth, for chambers with entirely white walls and with entirely black walls. In both cases, the detection probability decreases with increasing radius of the chamber. The optimum depth varies with radius, between around 1\,m to 2\,m depth for the white chamber of 1\,m to 3\,m  radius and around 1.5\,m to 2\,m depth for the black chamber. The depth for optimum detection efficiency tends to be lower than the depth of at least $5X_0$ required for good containment of electromagnetic shower particles (red dashed line in Fig.~\ref{fig:dim_scan_10MeV}). Moreover, our studies on double-layered WCD arrays show that a depth of at least 2.5\,m is required for efficient $\gamma$/hadron separation (see Section \ref{sec: gamma hadron}). 

The muon identification ability of the lower chamber is dependent on the upper chamber depth, since the upper chamber needs to efficiently shield the lower chamber from electromagnetic particles. This shielding effect is illustrated in Table ~\ref{table:peLowerSheilding}, which shows the mean charge in the entirely white lower chamber, for different upper chamber depths (3.8\,m diameter and fixed 0.7\,m lower chamber depth) and $\gamma$-rays of ~$\sim$100\,MeV  and ~$\sim$1\,GeV. As the upper chamber depth increases, the mean number of photo-electrons in the lower chamber is seen to decrease.

Based on these discussions on geometry, we use a double-layered WCD with dimensions fixed at 3.8\,m diameter and 2.5\,m upper chamber depth for studies on material properties of linings.  
\begin{table}[h!]
\centering
\begin{tabularx}{0.48\textwidth} { 
    >{\raggedright\arraybackslash}X 
   >{\centering\arraybackslash}X 
   >{\raggedleft\arraybackslash}X  }
 Upper Depth (m) &  lower (~$\overline{n_{pe}}~[100~\rm{MeV}]$) &  lower (~$\overline{n_{pe}}~[1~\rm{GeV}]$)  \\
 \hhline{===}
 1.7 & 13 & 78\\
 2.1 & 12 & 54 \\
 2.5 & 10 & 38 \\
 2.9 & 10 & 29 \\
 \hhline{===}
\end{tabularx}
\caption{The mean number of photo-electrons in the lower chamber of a WCD unit for $\gamma$-ray energy of ~$\sim$100\,MeV  and ~$\sim$1\,GeV for different upper chamber depths and fixed 0.7\,m lower chamber depth, for vertical $\gamma$-rays injected across the top surface. Upper chamber depth creates a shielding effect, as with increasing this depth the number of photo-electrons in the lower chamber decreases. We expect $\sim$~40 photo-electrons for 2 GeV muons for all these different upper chamber depths and fixed 0.7\,m lower chamber depth.}
\label{table:peLowerSheilding}
\end{table}

 Multiple scattering of electrons pair produced by the incoming gammas, or in later generations of a cascade, result in Cherenkov light emission in any direction in the tank chamber. The reflectivity of the walls affects the particle detection efficiency of the detector unit, as reflective walls result in isotropisation of the emitted light. In Fig. \ref{fig:dim_scan_10MeV}, we already see a comparison of a chamber with entirely white walls and one with entirely black walls. As expected, the increased photon path length and the nearly isotropic scattering of photons in the white chamber increases the probability of light collection compared to a black chamber. The white chamber provides detection  efficiencies of 70\% and more.

Entirely white walls will provide best efficiency, but will result in Cherenkov photon arrival times with long tails, as the decay time scale is governed by both the wall reflectivity and the water transparency; at least the latter may vary over time and between detectors. We consider combinations of white and black chamber walls in order to limit the number of `late' photons. 

\begin{figure}[ht]
\centering
\includegraphics[width=0.48\textwidth]{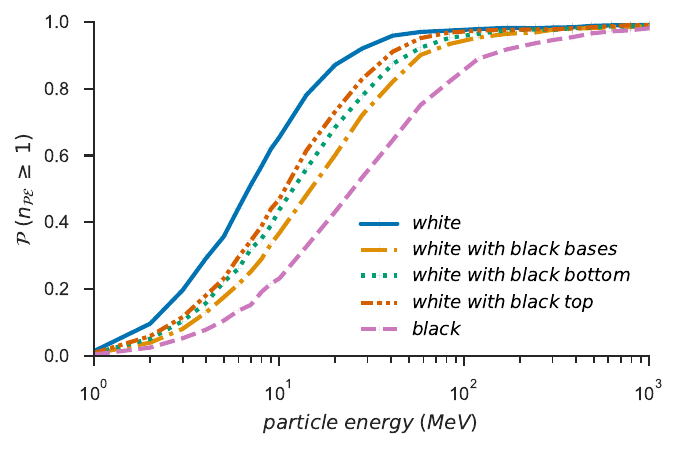}
\caption{Probability of detecting one or more photo-electrons for vertical $\gamma$-rays injected across the top of the upper chamber of the double-layered WCD (3.8 m diameter and 2.5 m depth) with an 8” PMT. Different curves correspond to different choices for the reflectivity of different internal surfaces of the chamber.
}
\label{fig:cDLWCD_comparison_sensitivity}
\end{figure}

We compare all-white and all-black chambers with chambers that have a black top, black bottom, or black bases (i.e. top and bottom), for vertical 1\,MeV to 1\,GeV $\gamma$-rays injected uniformly across the top surface, as shown in Fig.~\ref{fig:cDLWCD_comparison_sensitivity}. We see an improvement in the particle detection efficiency of at least partially white chambers over an entirely black chamber.

 Moreover, as the number of photo-electrons per unit energy scales like $\text{Area}_{\text{photo-cathode}}/\text{Area}_{\text{chamber}}$ , we show in Tab.~\ref{table:peperenergy} the average number of photo-electrons produced per 20 MeV of $\gamma$-ray energy, for vertical 1\,MeV to 100\,MeV $\gamma$-rays injected uniformly across the top surface. Here a 20 MeV $\gamma$-ray produces $\sim$1~pe in an entirely black chamber, and produces three times as many photo-electrons in an entirely white chamber. The differences at such low energies for different material combinations arise due to the effective production of a diffuse glow at the top of the tank which a white top helps to deflect more towards the PMT. To efficiently detect ${\le}20$\;MeV\;$\gamma$-rays would require reflective materials and/or more photo-cathode efficiency/area.

\begin{table}[h!]
\centering
\begin{tabularx}{0.48\textwidth} { 
    >{\raggedright\arraybackslash}X 
   >{\centering\arraybackslash}X 
   >{\raggedleft\arraybackslash}X  }
 Upper Chamber Material & ~pe~/~20\,MeV  \\
 \hhline{==}
 white & 2.9 \\
 white with black bases & 1.4  \\
 white with black bottom & 1.6  \\
 white with black top & 1.9 \\
 black  & 1.0 \\
 \hhline{==}
\end{tabularx}
\caption{The average number of photo-electrons produced per 20 MeV $\gamma$-ray energy, for vertical 1\,MeV to 100\,MeV $\gamma$-rays injected uniformly across the top surface in the upper chamber of a double-layered WCD unit (3.8\,m diameter and 2.5\,m depth) and an 8” PMT with different materials. }
\label{table:peperenergy}
\end{table}

\begin{figure}[!ht]
\centering
\includegraphics[width=0.48\textwidth]{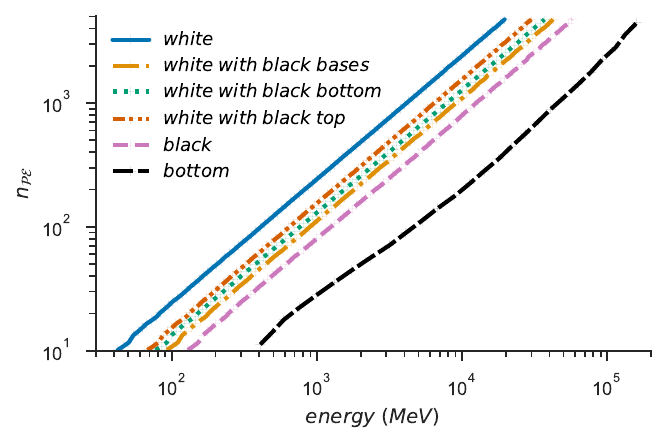}
\caption{The average number of photo-electrons as a function of electromagnetic energy for upper chambers of double-layered WCD units (3.8\,m diameter and 2.5\,m depth) with different materials. Also shown is the corresponding value for the lower chamber (3.8\,m diameter and 0.5\,m depth) as a dashed black line.}
\label{fig:comparison_ampres_cDLWCD}
\end{figure}

\begin{figure}[!ht]
\centering
\includegraphics[width=0.48\textwidth]{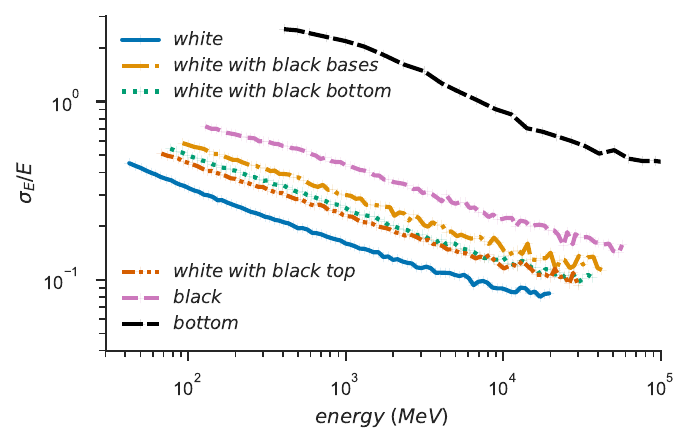}
\caption{Fractional rms energy resolution as a function of incident electromagnetic energy for the upper chamber of double-layered WCD units (3.8\,m diameter and 2.5\,m depth) with different internal surface reflectivities. The dashed black line shows the corresponding value for the lower chamber (3.8\,m diameter and 0.5\,m depth).}
\label{fig:comparison_res_cDLWCD}
\end{figure}

In order to illustrate the performance of the calorimetric measurement of shower particles, we show in Fig.~\ref{fig:comparison_ampres_cDLWCD}  the relation between electromagnetic energy and the number of photo-electrons and in Fig.~\ref{fig:comparison_res_cDLWCD} the resolution $\sigma_E/E$ as a function of electromagnetic energy, for different combinations of wall materials. As before, the all-white chamber behaves best in terms of pe yield and energy resolution, the all-black chamber is worst, while the mixed-wall chambers are intermediate. For incident particle energies above $\sim$500 MeV, the lower chamber also becomes sensitive to electromagnetic cascades, as shown in the same figure. Although the lower chamber provides rather poor resolution (Fig.~\ref{fig:comparison_res_cDLWCD}), 
it may help to extend the dynamic range of the system, which is important for detector units close to the shower core and/or in very high-energy showers.

The impact of the reflectivity of the upper chamber material on the angular resolution and $\gamma$/hadron separation is discussed in later sections (cf.~\ref{sec: gamma hadron},~\ref{sec: angular reconstruction}), and is found to be modest.

 Furthermore, we compare the performance of the double-layered WCDs with the performance of single-chamber WCDs similar to HAWC and LHAASO units that are currently in use. The four configurations are listed below.
 
\begin{enumerate}[label=(\Alph*)]
    \item A white double-layered WCD (3.8\,m diameter and 2.5\,m depth) with a black top and an 8-inch PMT
    \item A HAWC-like single-layered unit (7.3\,m diameter and 4\,m depth) with black walls, a central 10" PMT and three 8-inch PMTs
    \item  A LHAASO-like black unit (5\,m$\times$5\,m square, 4.5\,m depth) with an open top and an 8-inch PMT\footnote{An 8-inch PMT is used for comparison and does not reflect the actual PMT size(s) currently used in LHAASO WCD units}
    \item A white double-layered WCD unit with an alternative geometry (3.4\,m diameter and 3.0\,m depth) with a black top and an 8-inch PMT
\end{enumerate}

Configuration D is a deeper and thinner version of configuration A, both are double-layered WCDs. The other two are replications of the WCD units of the existing widefield observatories. The central PMT for configuration D is a higher quantum efficiency PMT similar to HAWC\citep{Abeysekara_2017}. 

\begin{figure}[!ht]
\centering
\includegraphics[width=0.48\textwidth]{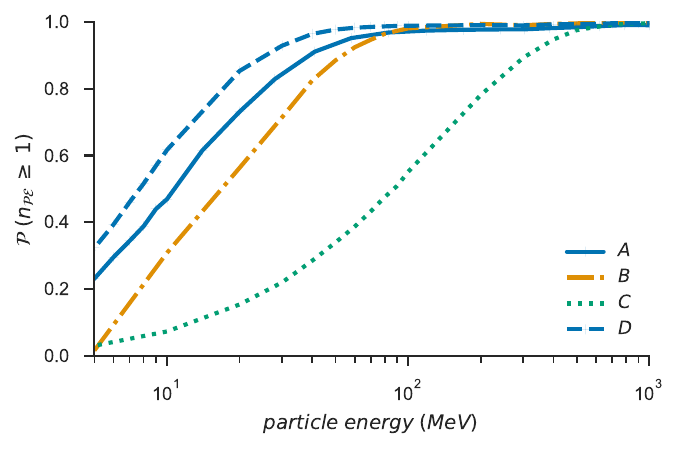}
\caption{Detection probability for vertical 5\,MeV to 1\,GeV $\gamma$-rays injected across the top surface of different WCD designs: (A) a cylindrical white double-layered WCD (3.8\,m diameter and 2.5\,m depth) with a black top and an 8-inch PMT; (B) a cylindrical single-layered HAWC-like unit (7.3\,m diameter and 4\,m depth) with black walls, a central 10" PMT and three 8-inch PMTs; (C) a LHAASO WCD-like  entirely black unit (5\,m $\times$ 5\,m square, 4.5\,m depth) with an open top and an 8-inch PMT; and (D) a white cylindrical double-layered WCD unit (3.4\,m diameter and 3.0\,m depth) with a black top and an 8-inch PMT.}
\label{fig:comparison_sensitivity}
\end{figure}

In order to investigate particle detection efficiency as a function of energy, we inject  vertical 5\,MeV to 1\,GeV $\gamma$-rays uniformly distributed across the top surface of these different configurations of WCDs.
Fig.~\ref{fig:comparison_sensitivity} shows a comparison of the detection probability. The upper chamber of the double-layered WCD (configurations A and D) has improved particle detection efficiency for $\gamma$-rays over both HAWC-like and LHAASO WCD-like designs, due to the reflective walls and -- in case of the LHAASO-like design -- the better ratio of PMT area to chamber surface. The HAWC and LHAASO WCD arrays employ non-reflective surfaces, which reduces the time spread of light reaching the photo-sensor (see Section \ref{sec: timing}), but results in a less-uniform response and reduced overall detection efficiency compared to reflective surfaces.

\subsection{Particle Arrival Time}
\label{sec: timing}

The arrival time distributions of Cherenkov photons at the PMT determine the time resolution that can (potentially) be achieved with a WCD unit; 
these time distributions and the background rates (Section~\ref{sec: noise rates}) also place important requirements on the readout electronics of a unit, including the length of the signal trace, buffering requirements, and trigger design. The aspect ratio and the material choice of the WCD units determine these time distributions and hence influence time resolution, which in turn impacts the achievable angular resolution for showers as discussed in Section~\ref{sec: angular reconstruction}. 

\begin{figure}[!ht]
\centering
\includegraphics[width=0.48\textwidth]{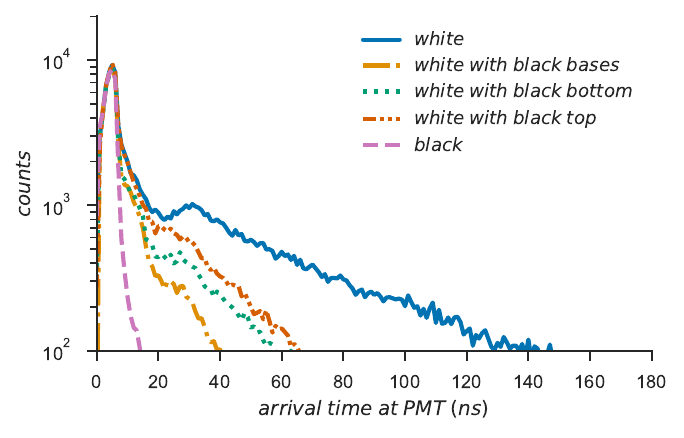}
\caption{Arrival time distribution of photons arriving at the PMT of the upper chamber of a double-layered WCD unit (3.8\,m diameter and 2.5\,m depth, similar to configuration A) with different materials, for vertical $\gamma$-rays at 80\,MeV. We note that the shape of these distributions depends only very weakly on the energy of the incident particle.
}
\label{fig:time_dist_r19x25r19x05_80MeV_gamma}
\end{figure}

In Figure \ref{fig:time_dist_r19x25r19x05_80MeV_gamma} we examine the arrival time of Cherenkov photon distributions resulting from 80\,MeV vertical $\gamma$-rays injected across the top surface of double-layered WCD units (3.8\,m diameter and 2.5\,m depth) with a combination of different material properties. The timing distributions show that a white walled chamber with one or both of the top and bottom surfaces black reduces the tail of the arrival distribution of photons by $\sim$30--40\% at this energy.

\begin{figure}[!ht]
\centering
\includegraphics[width=0.48\textwidth]{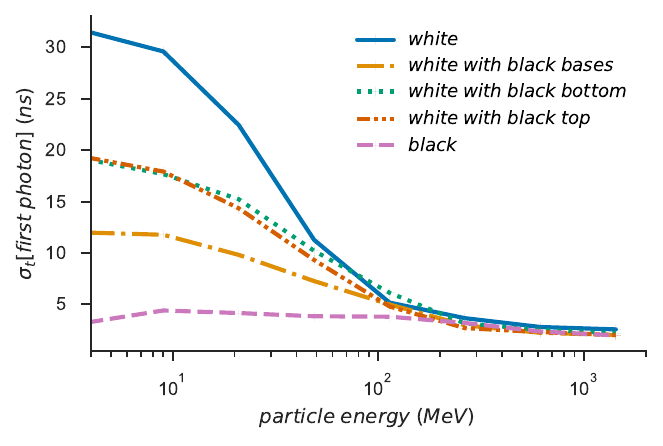}
\caption{Time resolution (rms) resulting from the arrival times of the first photo-electron as a function of incident $\gamma$-ray energy, for the upper chamber of a double-layered WCD unit (3.8\,m diameter and 2.5\,m depth) with different material properties and for vertical $\gamma$-rays.}
\label{fig:res_fpe_gamma}
\end{figure}

To study differences in time resolution, we examine the time of the first detected Cherenkov photon for each material combination (see Fig.~\ref{fig:res_fpe_gamma}). For black detectors, these first photons are dominated by direct light, hence the timing resolution (defined with respect to a particle entering the upper chamber) is mostly defined by the width of the detector. For (partially) white detectors, the timing resolution depends on the amount of Cherenkov light produced: low energy particles have a higher chance to be detected due to reflected light alone. While this adds more signals available to the reconstruction, these small signals show a worse time resolution. All material combinations show similar time resolution of the first photon above for particle energies above $\sim200$\,MeV because the probability to detect direct light becomes close to unity.


\begin{figure}[!ht]
\centering
\includegraphics[width=0.48\textwidth]{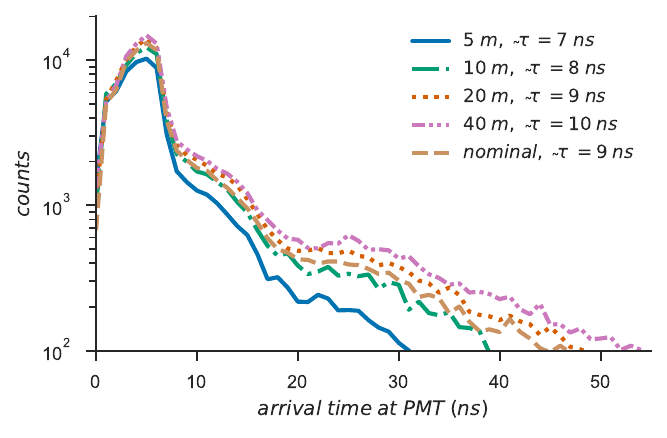}
\caption{Time distribution of photons arriving at PMT of the upper chamber of a double-layered WCD unit (3.8\,m diameter, 2.5\,m depth, and white walls with black bases) with water of varying absorption length, for vertical 120 MeV $\gamma$-rays injected across the top surface. The mean arrival time in each case is indicated in the figure legend.}
\label{fig:water_abs_len}
\end{figure}

Moreover, Fig.~\ref{fig:water_abs_len} illustrates the impact of changing water quality/absorption length on the arrival time distribution, for an upper chamber with white walls and black top and bottom. There is a very modest impact on the time distribution provided an absorption length of $>$10~m can be maintained. A nominal absorption length of 17\,m at 400\,nm is used in the simulations.

\begin{figure}[!ht]
\centering
\includegraphics[width=0.48\textwidth]{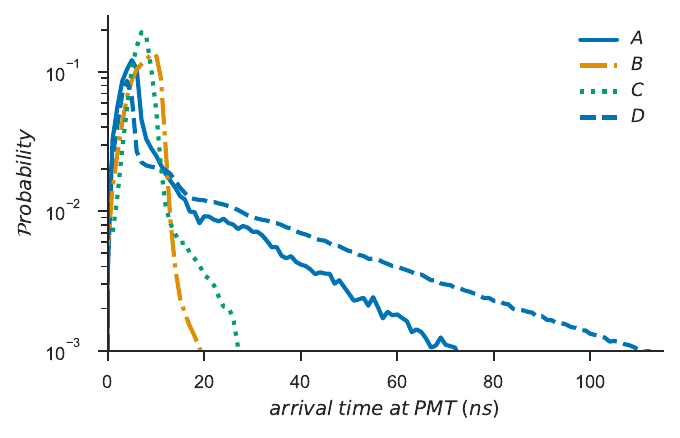}
\caption{Time distribution of photons arriving at the PMT for the four different WCD configurations from Section \ref{sec: sensitivity}: (A) a cylindrical white double-layered WCD (3.8\,m diameter and 2.5\,m depth) with a black top and an 8-inch PMT; (B) a cylindrical single-layered HAWC-like unit (7.3\,m diameter and 4\,m depth) with black walls, a central 10" PMT and three 8-inch PMTs; (C) a LHAASO WCD-like  entirely black unit (5\,m $\times$ 5\,m square, 4.5\,m depth) with an open top and an 8-inch PMT; and (D) a white cylindrical double-layered WCD unit (3.4\,m diameter and 3.0\,m depth) with a black top and an 8-inch PMT. The photons are initiated by vertical $\gamma$-rays at 80\,MeV. Only the the timing distributions from the central 10" PMT in the HAWC-like design and the PMT in the upper chamber of the double-layered designs are shown.
}
\label{fig:time_dist_comparison_designs_80MeV_gamma}
\end{figure}

Finally, we look at the four different WCD designs from Section \ref{sec: sensitivity}. Fig. \ref{fig:time_dist_comparison_designs_80MeV_gamma} shows Cherenkov photon arrival time distributions, resulting from 80\,MeV vertical $\gamma$-rays. As expected, reflective materials result in longer tails in the timing distribution of both double-layered designs compared to the HAWC and LHAASO-like designs which use entirely black walls. 

\subsection{Background WCD Trigger Rates}
\label{sec: noise rates}

Air shower arrays typically trigger based on a coincidence of triggers from individual units. The required minimal number of coincident units determines the energy threshold of the array, and depends both on the trigger rate of individual units and on the coincidence time window. The background trigger rate of WCD units will depend on their detection probability for different types of background particles, and hence on their geometry and wall materials; we expect chambers with some white surfaces to experience increased background rates with respect to entirely black chambers.

\begin{figure}[!ht]
\centering
\includegraphics[width=0.48\textwidth]{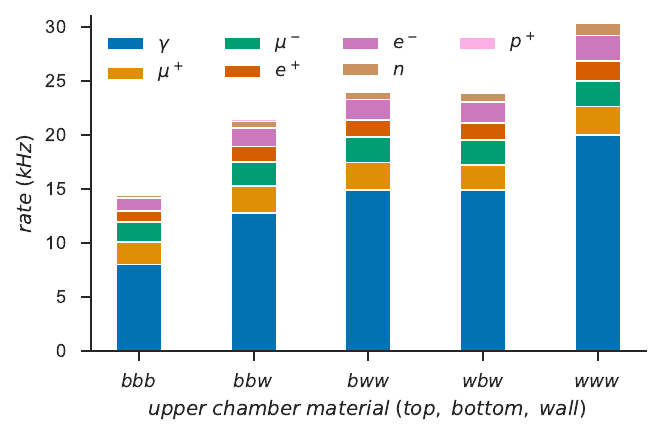}
\caption{Expected trigger rates at single pe\ threshold from incident particles for the upper chamber of a double-layered WCD unit (3.8\,m diameter and 2.5\,m depth) comprising an 8-inch PMT and various combination of materials (b = black, w = white), at an altitude of 4900 m. Contributions from $e^\pm$, $\mu^\pm$, $\gamma$, $p^+$ and $n$ are shown.}
\label{fig:backgrounds_cdlwcd_material}
\end{figure}

To investigate these rates, we use EXPACS/PARMA \citep{Sato_2016} to calculate terrestrial cosmic-ray fluxes and angular distributions of neutrons, protons, electrons, positrons, muons, anti-muons, and photons with energies ranging from $1$--$10^6$\,MeV at an altitude of 4900\,m\, a.s.l.\ and location 13\textdegree\,51'\,24"\,S, 71\textdegree\,1'\,30"\,W (a location that well characterizes the magnetic effect at all the sites under consideration for SWGO, see \citep{2022icrc.confE.689D} for an overview). To calculate the trigger rates, we inject particles aimed at a  hemisphere of 12\,m radius centered at the detector under test so that it can hit the detector from all sides (i.e 0 to 90 degree in zenith angle with a $sin \cdot cos$ distribution). Such a radius includes the edges of the surrounding detectors to account for scattered particles and shielding effects. In order to include shielding, a mini-array made up of 20 tanks with a separation of 0.12\,m between each tank was used. The expected single-pe\ trigger rates in the upper chamber of double-layered WCD units (3.8\,m diameter and 2.5\,m depth) composed of different materials are shown in Fig.~\ref{fig:backgrounds_cdlwcd_material}, broken down into contributions from different background particle species. Reflective walls increase the detection probability of low energy particles, resulting in an increased background rate. Nevertheless, we expect an array of white-walled detectors to provide the lowest energy threshold, once the coincidence level is adjusted to obtain a negligible rate of array-level noise triggers. Note that these background trigger rates do not include the single-pe thermal noise rate from the PMT itself, or possible contributions from radioactive decays in the water. 

These EXPACS/PARMA-based rates have been verified with CORSIKA air shower simulations to generate ground-level particle rates. Looking at different altitudes, the rates at 4900\,m are approximately 1.5 times the rates at an altitude of 4100\,m ($\sim$32\,kHz and $\sim$21\,kHz), at the same latitude and longitude, of an entirely white upper chamber of a double-layered WCD unit (3.8 m diameter and 2.5 m depth). 

The same method was used to predict the noise trigger rates of already operating HAWC tanks, which agree with the measurements reported by the HAWC collaboration. Trigger rates were predicted for the central 10'' PMT in a single tank in the HAWC central array and a single HAWC outrigger tank, at the actual HAWC altitude of 4100\,m. Along with an afterpulsing rate prediction, the rates amount to be $\sim$36\,kHz for a central PMT in a tank of the main HAWC array and $\sim$4.5\,kHz for a tank that is part of the outrigger array. Contributions from dark rates that varies around 1-3\,kHz with temperature, voltage and PMT dependence\citep{Hamamatsu_2019} are not included in the method. The afterpulsing rates were estimated with a first-order calculation that assumes an afterpulse probability of 0.02\% for the 10'' PMTs, as reported for these PMTs for a 2013 study for the Double Chooz experiment \citep{2013JInst...8P4029H}. The prediction can vary, as PMTs can have higher afterpulsing probabilities of 10\% if they are degraded, which would mean that the afterpulse rate is higher than the predicted amount for some of the tanks. Indeed, the HAWC data shows a large spread.  The central 10'' PMTs of HAWC were reported to have a hit rate of 40-50\,kHz \citep{Abeysekara_2017}. The outriggers of HAWC were reported to have rates around 4-8\,kHz \citep{2019ICRC...36..736M}, which agrees with our predictions with this method.

\section{Lower Chamber Optimisation}
\label{sec:lower}

In the dual-layer approach, muons are identified based on the signal from the lower chamber, or, more generally, by comparing the signals in the upper and lower chambers. Since the lower chamber is not used for timing, we assume white-walled lower chambers for optimal light yield. The depth of the lower chamber, combined with the photosensor area, determines the light yield. While in principle a few detected photons per muon are sufficient to tag muons with reasonable efficiency, our studies of algorithms for muon identification in showers (Section~\ref{sec: gamma hadron}) indicate that larger signals are desirable to reduce the number of misidentified muons. Here, we discuss how the muon signal depends on the geometry of the lower chamber.

\subsection{Depth and Reflectivity of the Lower Chamber}

\begin{figure}[!ht]
\centering
\includegraphics[width=0.48\textwidth]{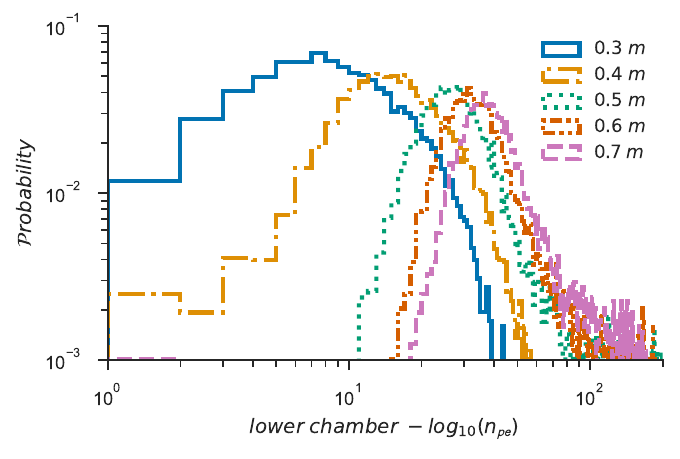}
\caption{The number of photo-electrons produced by vertical  2\,GeV muons in a WCD unit with an entirely white lower chamber, for  various different chamber depths.}
\label{fig:lower_chamber_npe_depth_2GeV}
\end{figure}

\begin{figure}[!ht]
\centering
\includegraphics[width=0.48\textwidth]{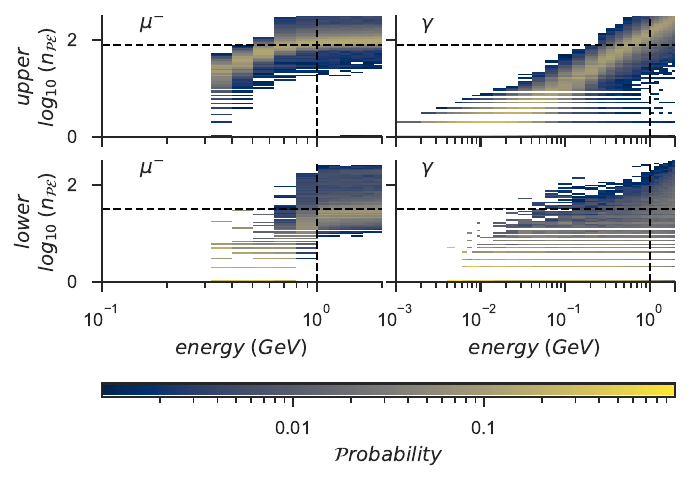}
\caption{The number of photo-electrons in a WCD unit with an entirely white upper (3.8 m diameter and
2.5 m depth) and lower (3.8 m diameter and
0.5 m depth) chamber. On the left, for $\mu^-$ with energy ranging from 100\,MeV to 2\,GeV; on the right, for $\gamma$-rays with energy ranging from 1\,MeV to 2\,GeV.  Dashed lines show 80\,pe\ and 1\,GeV in the upper plots and 32\,pe\ and 1\,GeV in the lower plots.}
\label{fig:compare_gamma_mu}
\end{figure}

We study lower chambers with white walls and with a PMT placed at the centre of the top of the chamber facing downwards. For best photon collection efficiency and hence minimal chamber depth, the PMT base protrudes into the upper chamber so that only the active cathode area of the PMT is visible in the lower chamber (see ~\ref{sec:LowerPMT:appendix}). The lower chamber has the same diameter as the upper to allow partitioning of a single cylindrical detector unit. A uniform response for muons above $\sim$600~MeV is expected since muons lose $\sim$2 MeV/cm in water (see lower chamber in Fig.~\ref{fig:compare_gamma_mu}). 

Figure \ref{fig:lower_chamber_npe_depth_2GeV} shows that for depths less than  $\sim$0.5\,m, the muon signal has a much larger spread compared to depths greater or equal to  $\sim$0.5\,m. Moreover, for depths greater than $\sim$0.5\,m, the muon signal is reliably $>$10 photo-electrons for $>$2 GeV muons, and the signal increases roughly proportionally to the track length of the impinging particle as expected. It is seen that depths greater or equal to $\sim$0.5\,m would give reliable muon signals, and since a lower chamber that is smaller in depth would minimize costs (although most of the cost is driven by the upper chamber), the studies here suggest that the depth for the lower chamber should be at least $\sim$0.5\,m.

To explore the separation power of such a unit, we compare vertical $\gamma$-rays with energies from 1\,MeV to 2\,GeV in an entirely white WCD unit (3.8 m diameter, 2.5 m upper and 0.5 m lower depth) with vertical $\mu^{-}$ with energies from 100\,MeV to 2\,GeV in an identical WCD unit (see Fig.~\ref{fig:compare_gamma_mu}).  
Muon identification is possible as the number of photo-electrons detected in the two chambers is constant above some energy threshold and remains fairly stable until very high energies, where effects such as bremsstrahlung will need to be considered. Additionally, the peak of the muon energy distribution in air showers is around 2--3\,GeV~\citep{Schoorlemmer_2019}. The ratio of photo-electrons in the two chambers will enable muon selection on a tank-by-tank basis. Other shower information such as the location of the shower core relative to the detector unit and the number of units hit is also beneficial for muon identification.

\section{Array Simulations}
\label{sec:array}

To relate the performance of individual WCD units to the performance of an array as a whole and in particular to the achievable angular resolution and $\gamma$/hadron separation power, we carried out simulations based on the SWGO reference design \citep{SWGO_2021}. We continue using the \emph{HAWCSim} tool that has been validated by the HAWC collaboration, as explained in Section \ref{sec: simulation overview}. We simulate an inner array with a high fill factor of $\sim$80\% spread over $\approx$80,000\,m$^2$ and a sparser outer array with a fill factor of 8\% spread over $\approx$220,000\,m$^2$, placed at high elevation (4900\,m.a.s.l). The dense inner array serves to enhance the sensitivity for low to mid energy $\gamma$-rays and also to increase the muon sensitive area. The sparser outer array is designed for a large collection area at the highest energies, but will not be used in the analysis presented here. Comparisons with the currently operating HAWC and LHAASO arrays are also beyond the scope of this paper, since the layouts, altitudes and the algorithms used in the analysis are different for these WCD arrays.

\subsection{Estimating Effective Area}
\label{sec: array background rates}

\begin{figure}[!ht]
\centering
\includegraphics[width=0.48\textwidth]{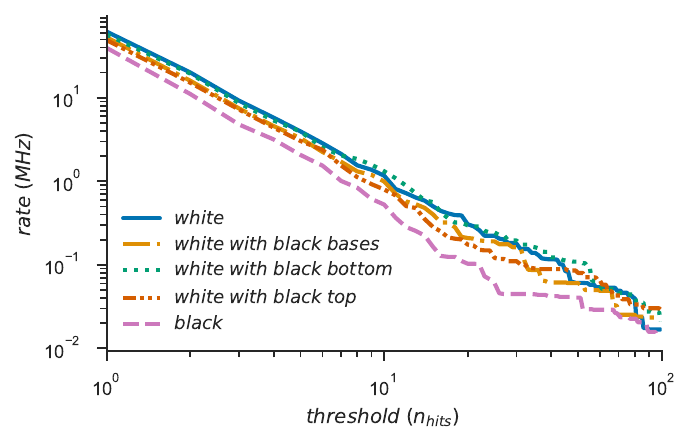}
\caption{Estimated cosmic ray array-level trigger rate for an array of double-layered WCDs (3.8\,m diameter, 2.5\,m upper depth, and 8-inch PMT) with different surface reflectivity choices, as a function of the WCD multiplicity ($n_{\rm hits}$) adopted for the array trigger decision.}
\label{fig:background_array_rate}
\end{figure}

To be able to compare the $\gamma$-ray induced air shower effective area for different material combinations, we first simulate proton (background) induced showers with a spectral index of -2, an energy range from 0.001--30 TeV and zenith angle from $0^\circ \; {\rm to} \; 60^\circ$, at an altitude of 4900\,m.a.s.l. and shower core scattered over a radius of 2.5 km. These simulations are then weighted by the cosmic-ray flux~\citep{Lipari_2019} to reproduce the correct spectrum. As the core range is large compared to the detector size, we can further integrate to obtain approximate array trigger rates at different hit thresholds (see Fig.~\ref{fig:background_array_rate}). 

\begin{figure}[!ht]
\centering
\includegraphics[width=0.48\textwidth]{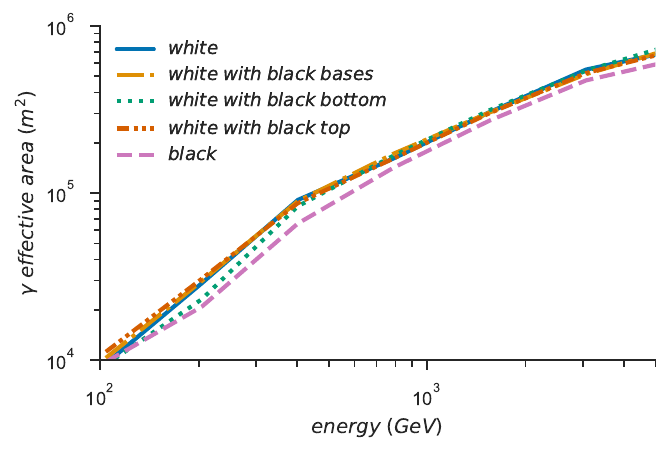}
\caption{Effective collection area for $\gamma$-ray initiated air showers for an array of double-layered WCDs (3.8\,m diameter, 2.5\,m upper depth, and 8-inch PMT) with different surface properties for the upper chamber.}
\label{fig:gamma_effective_area}
\end{figure}

We then derive the $\gamma$-ray effective area as a function of energy, as shown in Fig.\ref{fig:gamma_effective_area}. We first simulate $\gamma$-ray induced air showers with a spectral index of -2 for an energy range of 0.1--5 TeV, zenith angle from $0^\circ \; {\rm to} \; 60^\circ$, at an altitude of 4900\,m.a.s.l.\ and shower core scattered over a radius of 2.5 km with respect to the array center. From the proton simulations, we derive the required threshold in the number of array hits, such that the array trigger rate is 100 kHz (ensuring at least a few 10s of tanks trigger). With this threshold, we use the $\gamma$-ray simulations to derive the $\gamma$-ray effective area as a function of energy (see Fig.~\ref{fig:gamma_effective_area}). The simulations show that double-layered WCDs (3.8\,m diameter, 2.5\,m upper depth, and 8” PMT) with some reflective surfaces for the upper chamber provide slightly higher effective areas at low energies, compared to entirely black upper chambers. It should be noted however that more statistics, including hit timing information and large core range would be necessary to obtain realistic rates.

\subsection{Identifying Hadronic Showers}
\label{sec: gamma hadron}

The double-layered WCD design provides an alternate way of vetoing hadron-induced showers compared to single-chambered WCD designs. The critical parameter that we will explore in the following is the depth of the upper chamber, that determines the level of electromagnetic punch-through into the lower chamber (see Section \ref{sec:upper} for the influence of upper chamber depth on single WCD unit performance).

\begin{figure}[!ht]
\centering
\includegraphics[width=0.48\textwidth]{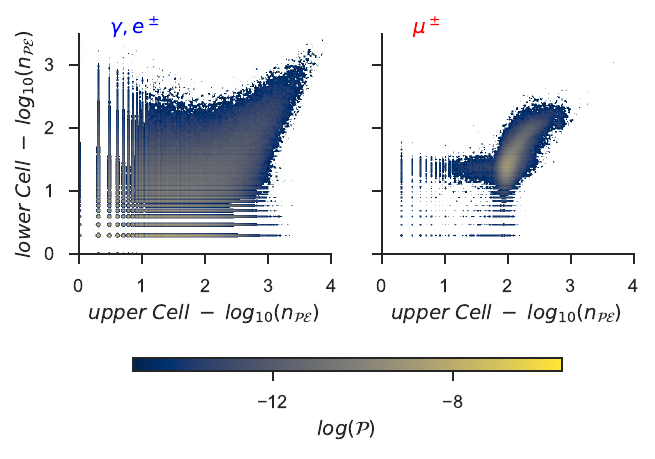}
\caption{Distribution of signals in the upper and lower chambers of entirely white double-layered WCDs from vertical proton shower simulations, excluding the region within 20\,m of the shower core. These distributions are used as templates for the likelihood-based background separation. }
\label{fig:muonTemplate}
\end{figure}

For this purpose we implement a template-based maximum log-likelihood method to discriminate between $\gamma$-ray and hadron-induced air showers, similar to template-based reconstruction methods by the HAWC Collaboration \citep{Joshi_2019}. Given a known core location and air shower direction, we generate templates for the charge in the upper and lower chambers in an array of double-layered WCDs, similar to the distribution shown in Fig.~\ref{fig:muonTemplate}, for vertical air showers. We generate separate templates for $\mu^\pm$ and for other charged particles. Next, we test simulation events, where we assign a likelihood value based on the charge deposited in the two chambers for each secondary particle impinging on an individual tank. Subsequently, we calculate a Likelihood Ratio (LR) to tag those particles with a $LR > 0$ as candidate muons and compare them to the Monte-Carlo truth.

\begin{figure}[!ht]
\centering
\includegraphics[width=0.48\textwidth]{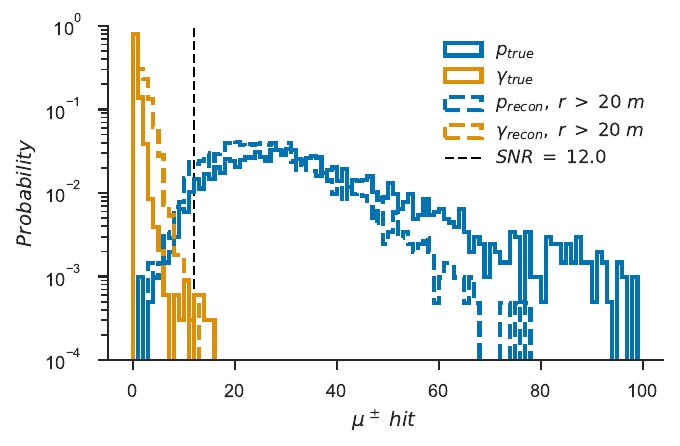}
\caption{Probability of detecting muons in detector units more than 20\,m from the shower, for $\gamma$-ray- and hadron-initiated showers detected with entirely white double-layered WCDs in the dense inner array of the SWGO Reference Configuration. Showers are simulated with a core located at the centre of the array. The black dashed line indicates the cut value giving the highest signal to noise ratio  ($\epsilon_{\gamma}/ \sqrt{\epsilon_{\text{proton}}}$) is maximum.}
\label{fig:gamma_hadron_sep}
\end{figure}

This muon tagging ability directly translates into $\gamma$/hadron separation efficiency, given the relative abundance of muons in hadron-initiated showers. Using the number of detector units hit as a proxy for the shower's energy,  we can subsequently distinguish $\gamma$-ray and hadron-initiated showers from the difference in the number of identified muons, as shown in Fig.~\ref{fig:gamma_hadron_sep} for vertical 2\,TeV $\gamma$-ray and 5\,TeV proton induced showers. The proton shower energy is chosen such the the average number of WCD hits is very similar for the two sets of events. To reduce misidentification due to punch-through of electromagnetic particles from the upper to lower chamber close to the shower core, we exclude WCDs within 20\,m from the core.

From the known core distance and air shower direction, we generate templates of charge in the upper and lower chambers in an array of entirely white double-layered WCDs (3.8\,m diameter and 2.5\,m upper chamber depth and 0.7\,m lower chamber depth \footnote{As we aim to optimise the upper chamber depth first, the lower chamber is left at a depth of 0.7 m and later optimised to 0.5 m with careful positioning of the lower PMT as discussed in \ref{sec:LowerPMT:appendix} }) for vertical 5\,TeV proton-induced air showers. 

After tagging different particle species for  2\,TeV $\gamma$-ray and 5\,TeV proton induced showers, we find a $\gamma$/hadron separation efficiency while varying the upper chamber depth. The signal-to-noise-ratio (SNR) is $\epsilon_{\gamma}/ \sqrt{\epsilon_{\text{proton}}}$, where $\epsilon$ is efficiency. For upper chamber depths greater than 2.5\,m, the SNR plateaus for air showers at 0$^{\circ}$ zenith angle. Inclined showers are also expected to plateau, but as individual secondary particles from inclined showers can penetrate through the sides and multiple tanks, there is a decrease in overall SNR. Including the neighbouring tanks would mitigate the decrease in SNR for such inclined showers. Increasing detector unit radii or increasing the fill factor could also be beneficial for shielding from side-penetrating particles; one could also consider filling up the space between tanks with ground material as absorber. However, a complete study of inclined showers is beyond the scope of this paper. 

\begin{figure}[!ht]
\centering
\includegraphics[width=0.48\textwidth]{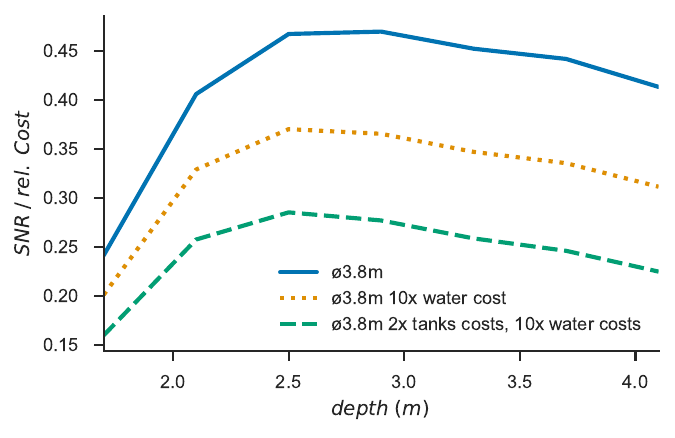}
\caption{Ratio of $\gamma$/hadron separation SNR to nominal cost in an array with entirely white double-layered WCDs with varying upper chamber depth, given different assumptions of the relative cost of water to photosensors and tanks, assuming a fixed number of tanks. Each WCD unit has a diameter of 3.8\,m and a fixed lower chamber depth of 0.7\,m.}
\label{fig:ghsepefficiencycost}
\end{figure}

We calculate the SNR-to-cost ratio by attempting to account for the cost scalings for tanks with depth, volume of water, and the fixed cost of photosensors. We find that for vertical showers the SNR-to-cost ratio peaks at $\sim$2.5\,m, and this depth is relatively insensitive to the cost assumptions made (see Fig.~\ref{fig:ghsepefficiencycost}). Based on these findings, the upper chamber depth of 2.5\,m has been used in the comparisons of Section \ref{sec:upper} and while investigating different chamber materials. 

\begin{figure}[!ht]
\centering
\includegraphics[width=0.48\textwidth]{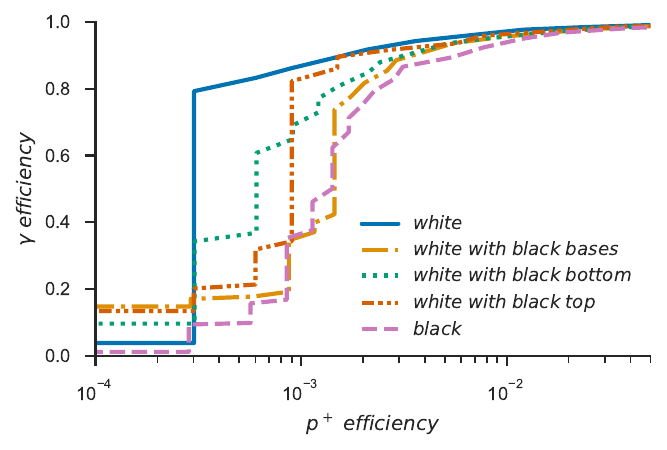}
\caption{$\gamma$/hadron separation efficiency for an array of double-layered WCDs (3.8\,m diameter and 2.5\,m depth) with $\sim$80\% fill factor varying material reflectivity, using an exclusion region of 40\,m around the core location for reconstructed Gamma showers of 2 to 7 TeV.}
\label{fig:ghsep_efficiency}
\end{figure}

To show the effect of various materials in the upper chamber for double-layered WCDs (3.8\,m diameter, 2.5\,m upper depth, 0.5\,m lower depth, with 8-inch PMTs), we again implement a template-based maximum likelihood method to discriminate between $\gamma$-ray and hadron induced air showers. We simulate an ensemble of vertical $\gamma$-ray and proton initiated showers following an E$^{-2}$ spectrum up to 100\,TeV energy, with shower core location at the centre of the array. Fig.~\ref{fig:ghsep_efficiency} shows the derived $\gamma$/hadron separation efficiency (excluding a 40\,m  region around the core) for different material combinations. We see that a background rejection power of $\sim$10$^{3}$ can be achieved at reconstructed energies (a simple model taking into account only the total number of photo-electrons seen in all of the upper chambers of the array) of a few TeV (2- 7 TeV) with high $\gamma$-ray efficiency. The separation power improves in all cases with one or more white surface(s). 

From Fig.~\ref{fig:ghsep_efficiency} we find a background efficiency of $3\times10^{-4}$ keeping good gamma efficiency for entirely white double-layered WCDs (3.8\,m diameter, 2.5\,m upper depth, 0.5\,m lower depth, with 8-inch PMTs). At similar energies this is at least a factor $\sim30$ improvement in rejection power compared to HAWC \citep{Abeysekara_2017} and LHAASO \citep{2021ChPhC..45h5002A}. This improvement factor is extremely promising despite the somewhat idealised nature of the simulations (showers only from zenith, at the array centre, and neglecting uncorrelated noise hits) as the performance in the final array would likely be significantly improved by the inclusion of additional parameters or more sophisticated treatment of the muon-based rejection.


\subsection{Angular Reconstruction}
\label{sec: angular reconstruction}

To reconstruct the direction of the air shower, we use the time of the first photon in each upper WCD without applying transition time spread in the PMT, or electronics time resolution. We first obtain a guess of the direction from a least-squares algorithm using the time difference of arrival of the photons between detector units.

We obtain the final best-fit direction via a likelihood fit to the arrival times in each hit detector unit, described in ~\citep{Hofmann_2020}. Parameterized distribution functions describe the distribution in energy, core distance and arrival time of shower particles at ground. We obtain PDFs for the time distribution by fitting Landau distributions to the arrival times binned as function of distance to the shower core and total charge, to obtain mean and width parameters for each bin ~\citep{Hofmann_2020}. A shower arrival time, $t_0$, is defined as the absolute arrival time of the first electromagnetic particle in the plane perpendicular to the shower axis. To separate the effects of the direction (i.e. timing) fit and the core position fit, and to minimise additional complexity due to array edges, we generate vertical $\gamma$-rays impacting at the centre of the array, and assume the core position to be known (as in practice the core location can be very precisely determined for well-contained events).

\begin{figure}[!ht]
\centering
\includegraphics[width=0.48\textwidth]{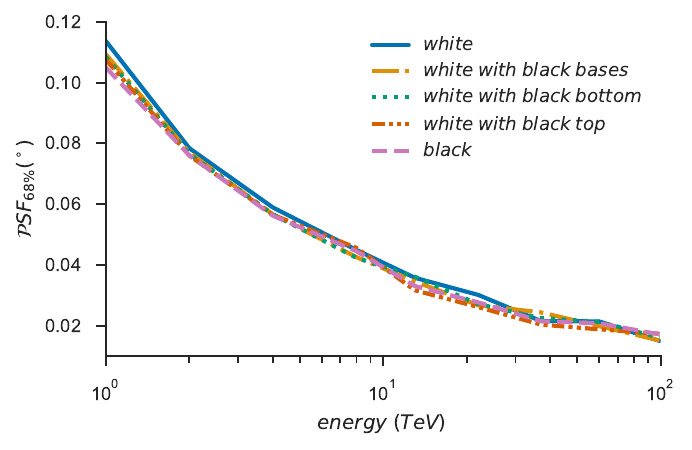}
\caption{Angular resolution for 1--100\,TeV vertical $\gamma$-rays at the centre of an array of double-layered WCDs (3.8\,m diameter and 2.5\,m depth) with $\sim$80\% fill factor -- for a range of options on surface reflectivity.}
\label{fig:angular_res_energy}
\end{figure}

We perform a three-parameter (time, offset and direction) likelihood fit (using MINUIT~\citep{James_1994}) to obtain the reconstructed shower directions and hence determine the angular resolution (see Fig.~\ref{fig:angular_res_energy}). As expected, the angular resolution improves with increasing energy. Differences between different wall options are very modest. The additional late-arriving Cherenkov photons that are a feature of white-walled chambers do not deteriorate the angular resolution as they are properly accounted for in the likelihood functions used in the fit; these likelihood functions are adapted for each configuration. 

Whilst this result represents a very idealised case, with showers from zenith landing at the centre of the array, with perfect timing resolution and no noise, it nonetheless illustrates that such arrays can potentially achieve an angular resolution much better than existing instruments of this type, regardless of the surface reflectivity of individual detector units ~\citep{Aharonian_2021, Abeysekara_2017}. Comparisons with different fill factors and  reconstruction algorithms are beyond the scope of this paper. 

\section{Summary \& Outlook}
\label{sec: summary}

We have studied a double-layered WCD array, aiming to improve both the energy threshold and the $\gamma$/hadron separation in comparison to LHAASO and  HAWC-like designs. Each detector unit in the array comprises two chambers with black or diffuse reflective wall linings and a PMT in each chamber. The upper PMT facing upwards is intended for timing and energy determination, while the lower PMT facing downwards will enable muon tagging, and provide the primary mechanism for $\gamma$/hadron separation.

The double-layer design is promising in terms of background rejection power, achieving a rejection power of $\sim$10$^{3}$ at energies of a few TeV, with high $\gamma$-ray efficiency. Investigating different options for chamber aspect ratio (depth-to-diameter) and reflectivity shows that a compact and highly reflective upper chamber lowers the energy threshold for the impinging particles and, in conjunction with a lower chamber, benefits $\gamma$/hadron separation. There is an increase in the tail in the timing distributions with one or more white surface in the upper chamber, however, these late-arriving Cherenkov photons do not deteriorate the angular resolution of the array when suitable reconstruction methods are employed. A partially reflective upper chamber is strongly motivated by these benefits in threshold and background rejection, with no negative impact on angular resolution.

Although the final optimization of a unit detector for SWGO requires finalized analysis algorithms and life cycle cost estimates that are beyond the scope of this paper, these studies show that a $\sim$3\,m deep, $\sim$4\,m diameter double-layered tank with some reflective material in the upper chamber is a promising option in terms of performance and cost-effectiveness. For the reference SWGO design where the diameter is fixed to 3.8\,m, an upper chamber depth of $\sim$2.5\,m  maximizes the SNR-to-cost ratio and provides sufficient shielding for the lower section, meanwhile a lower chamber depth of $\sim$0.5\,m is the minimum required for reliable muon signals.

 In the idealized case of vertical showers at the centre of the array, we obtained an angular resolution of several arc-minutes at 10 TeV and found that background rejection power of $\sim10^3$ is achievable. Our studies show that a densely packed ($\geq 80 \%$ fill factor), high altitude ($\sim5000$ m.a.s.l) array of double-layered WCDs has the potential to achieve superior angular resolution, reduce energy threshold and improve $\gamma$/hadron separation over existing WCD arrays.

\section{Acknowledgements}
We thank our colleagues within the SWGO collaboration for many helpful discussions and the use of the common shared software framework. We thank the HAWC collaboration for providing the AERIE software used here.

\appendix

\section{Note on positioning of the lower chamber PMT} 
\label{sec:LowerPMT:appendix}

To minimise the depth of the lower chamber, it suffices to only have the active cathode area of the PMT visible in the lower chamber. We optimise the PMT positioning in the lower chamber by pushing the base of the downward facing PMT in the lower chamber up into the upper chamber. To investigate the performance of this PMT positioning, we inject vertical 2\,GeV muons uniformly across the top surface of a double-layered WCD unit (3.8 m diameter, 2.5 m upper and 0.5 m lower depth) with an upward facing 8-inch PMT centered at the bottom of the upper chamber and a downward facing 8-inch PMT centered at the top of the entirely white lower chamber. We find that this PMT adjustment with the base of the downward facing PMT protruding ($\approx~10$cm) into the upper chamber results in a~x1.6 fold increase in the mean light yield compared to the entire PMT plus base in the lower chamber.

\bibliographystyle{elsarticle-num-names}  
\bibliography{dlwcd_bibs}





\end{document}